# Hall Anomaly and Vortex-Lattice Melting in Superconducting Single Crystal $YBa_2Cu_3O_{7-\delta}$

G. D'Anna[1], V. Berseth[1], L. Forró[1], A. Erb[2], and E. Walker[2]

[1]Institut de Génie Atomique, Département de Physique, Ecole Polytechnique Fédérale de Lausanne, CH-1015 Lausanne, Switzerland

[2]Département de Physique de la Matière Condensée, Univeristé de Genève, CH-1211 Genève, Switzerland

Abstract

Sub-nanovolt resolution longitudinal and Hall voltages are measured in an ultra pure $YBa_2Cu_3O_{7-\delta}$ single crystal. The Hall anomaly and the first-order vortex-lattice melting transition are observed simultaneously. Changes in the dynamic behavior of the vortex solid and liquid are correlated with features of the Hall conductivity $\sigma_{xy}$. With the magnetic field oriented at an angle from the twin-boundaries, the Hall conductivity sharply decreases toward large negative values at the vortex-lattice melting transition.

PACS numbers: 74.25.Fy, 74.60.Ge, 74.72.Bk

A type-II superconductor in the mixed state is penetrated by an array of magnetic vortices each of which contains a quantized amount of magnetic flux $\phi_o$. In a transport current there will be a Lorentz force density $\vec{F}_L = \vec{j} \times \vec{B}$ on the vortices, where $\vec{j}$ is the transport current density and $\vec{B}$ is the average vortex density. The vortices may be at rest, for example, if the Lorentz force is balanced by pinning forces or by finite size barrier effects, and the current flows without dissipation. The vortices may move with a steady mean velocity $\vec{V}$, and according to the Josephson relation $\vec{E} = -\vec{V} \times \vec{B}$ an electric field will appear. The Hall angle $\theta_H$ between the direction of the vortex motion and the Lorentz force is given by $\rho_{xy}/\rho_{xx} = \tan\theta_H$, where $\rho_{xx} = E_x/j$ is the longitudinal resistivity and $\rho_{xy} = E_y/j$ the Hall resistivity (we assume $\vec{B} = B\hat{z}$, $\vec{j} = j\hat{x}$). According to the classical models[1] the moving vortices should generate a Hall voltage with the same sign as observed in the normal state. However, in various high- and low-temperature type-II superconductors a Hall effect sign reversal has been reported by many authors, followed sometimes by a second sign change[2]. This "Hall anomaly" has attracted much theoretical attention[3-9], in particular since it has become clear that it is a general problem of vortex dynamics, not caused by some extrinsic material inhomogeneity.

Microscopically the trajectory of a single vortex in a transport current is determined by hydrodynamic and core forces which arise from the interaction with the superfluid and localized and delocalized excitations. A legitimate question is then: *does the first-order vortex-lattice melting transition, which is a hallmark of vortex dynamics in high temperature superconductors, produce an effect on the Hall behavior?* The question is relevant, in particular for the high quality $YBa_2Cu_3O_{7-\delta}$ (YBCO) single crystals of interest here[10,11], since theoretical analysis has shown that most of the latent heat observed at the vortex-lattice melting transition comes from changes in the entropy at microscopic length scales, and not solely from the change in the entropy associated with the vortex configuration[12-15].

Unfortunately, experimental difficulties to observe the Hall effect below the vortex-lattice melting transition exist. The most serious one is that in usual YBCO crystals, transport Hall experiments only probe the vortex-liquid phase because in the vortex-solid the critical current steeply increases. Pulsed current techniques[16] providing extremely high current densities have been used with success in YBCO films to detect the Hall voltage at low temperature, but have not been used to investigate the Hall behavior around the melting transition in clean crystals. Therefore, it is highly desirable to observe the Hall anomaly in a situation in which the vortex phases are clearly identified. In this Letter we exploit precise longitudinal and Hall resistivity data obtained in very pure YBCO crystals. Voltage resolution of 0.1nV and low critical currents permit us to determine the Hall conductivity below the vortex-lattice melting transition, and to clearly correlate the various vortex phases to features of the Hall behavior.

The crystal used here was grown in a $BaZrO_3$ crucible as described in ref.[17]. The crystal purity is the highest achieved to date and is similar to the ones used for specific heat experiments and scanning tunneling microscopy[11,18]. The micro-twinned crystal has dimensions 0.9x0.4 $mm^2$ in the a-b plane, and thickness 24 µm in the c-direction. The major twin family is at 45° from the long edge of the sample. Some untwinned domains and some twins at 90° from the dominant family are also present. The sample displays a sharp resistive transition at about $T_c$=93.5 K. Its oxygen content is estimated to be 6.94, and previous magneto-optical studies confirm its homogeneity[19]. A multi-contact configuration (seven contacts are used in the present experiment) was formed by evaporating gold dots on the upper surface and subsequent annealing for 45 hours at 400 °C in oxygen atmosphere. The longitudinal resistivity $\rho_{xx}$ and Hall resistivity $\rho_{xy}$ are measured simultaneously by injecting an AC current (30 Hz), sometimes on the top of a DC current, along the longest dimension of the crystal, and by measuring the in-phase voltages parallel and perpendicular to the current at the AC frequency. In particular, the Hall voltage is measured by removing the residual contact misalignment effect by taking the antisymmetric part of the signal, with respect to the magnetic field, from an already balanced three contact configuration according to a standard method (see for example ref. [20]). Moreover, in order to cancel out spurious voltages due to the motion of the voltage lead wires, we also reversed the current direction. This precaution is found to be necessary when we use large DC currents in regions where the true Hall voltage is very small, typically in the vortex-solid phase. We achieve a resolution of about 0.1 nV for each contact pair, and we measure the Hall voltage using current densities as low as $0.2 A/cm^2$. From the longitudinal and Hall resisitvities we determine the Hall angle $\theta_H$ ($\tan\theta_H = \rho_{xy}/\rho_{xx}$) and the Hall conductivity $\sigma_{xy} \approx \rho_{xy}/\rho_{xx}^2$.

Figure 1(a) displays $\rho_{xx}$ and $\rho_{xy}$ measured at 2 T for different AC and DC currents, as a function of the temperature T. The magnetic field is applied at α=4° from the c-axis. The alignment to the c-axis is accurate within ±0.1° as determined by searching the extrema of the longitudinal voltage[21]. By orienting the field at α=4° from the c-axis, the effect of twin-boundaries on vortex dynamics is strongly reduced, as discussed below. The lower panel, Fig. 1(b), displays the corresponding Hall angle $\theta_H$ and Hall conductivity $\sigma_{xy}$.

The $\rho_{xx}$ versus T curves in Fig. 1(a), and in the inset of Fig. 1(a), provide the kind of transport measurements by which one usually identifies the vortex-lattice melting transition. For small applied AC currents and zero DC current, the sharp drop in resistivity or the "kink" in the $\rho_{xx}$ curve has been associated with the first-order vortex-lattice melting transition[22]. More precisely, at very low current density $\rho_{xx}$ goes to zero when a superconducting percolative path of pinned vortex-solid develops across the sample[23]. We mark the percolation[24] temperature by $T_m$ in the figure, and we will call "vortex-liquid" the mixed state at temperatures $T>T_m$ and "vortex-solid" below $T_m$. By superimposing the AC current on top of a large DC current, a longitudinal resistivity different from zero is observed below $T_m$, showing that the vortex-solid is moving under the effect of the large Lorentz force. At the same time a non-zero Hall resistivity is detected, as shown in Fig. 1(a). In order to achieve a sufficient signal to noise ratio for the Hall voltage, we have used relatively large AC current densities as well, but the results are similar for smaller AC current densities. In conclusion, the $\rho_{xx}$ versus T data, and similar results obtained as a function of the magnetic field at constant temperature, provide us with a clear identification of the vortex phases, on the basis of which we can now discuss the Hall data.

By reducing the temperature from the normal state, namely in the vortex-liquid phase, the Hall resistivity $\rho_{xy}$ becomes negative (the Hall anomaly), together with the Hall angle $\theta_H$ and the Hall conductivity $\sigma_{xy}$, at a temperature denoted $T_x$, in Fig 1(b). By approaching the vortex-lattice melting transition using low current densities, near $T_m$ the Hall angle and the Hall conductivity can no longer be calculated, since both the longitudinal and Hall voltages are below our sensitivity, so the calculated $\theta_H$ and $\sigma_{xy}$ points are scattered. For this reason we simply end the curves when the large scattering starts in Fig. 1(b). On the other hand, by approaching the vortex-lattice melting transition using large current densities, the vortex-solid can be set into motion and the logitudinal and Hall voltages are clearly detected. In this case we observe a surprising Hall behavior. The Hall conductivity deviates from its behavior in the vortex-liquid phase and goes rapidly towards large negative values. Similar results are shown for different fields in the inset of Fig. 1(b). *This is the first experimental evidence that the vortex-lattice melting transition affects the Hall behavior*. It is also interesting to notice that the Hall angle and the Hall conductivity become much more noisy in the vortex-solid phase compared to the vortex-liquid. This noise is intrinsic to the vortex-solid and originates from the motion of bundles or channels of correlated vortices[25].

Figure 2 displays $\rho_{xx}$ and $\sigma_{xy}$ measured at T=88 K for different AC and DC currents, as a function of the magnetic field B. The field is applied at $\alpha=4°$ and at $\alpha=0°$ from the c-axis. First we discuss the data at $\alpha=4°$. The field $B_m$, where the longitudinal resistivity $\rho_{xx}$ goes to zero for the small applied AC current density and zero DC current, is our vortex-lattice melting transition. By superimposing the AC current on top of a large DC current, the vortex-solid below $B_m$ is set into motion. As in the previous measurements as a function of the temperature, *the Hall conductivity is found to decrease rapidly toward large negative values below* $B_m$. Similar results are shown in the inset of Fig. 2 for different temperatures in the form of the Hall force coefficient $\sigma_{xy}B$ [3-9].

The $\rho_{xx}$ data in Fig. 2 obtained with the field applied parallel to the c-axis, i.e. at $\alpha=0°$, show the strong effect of twin-boundaries on vortex motion. For fields below 3.4 T (denoted $B_{TB}$ in the figure), the longitudinal resistivity $\rho_{xx}$ is reduced with respect to the $\alpha=4°$ data, indicating that twin-pinning is much more effective for this orientation. Below about 2T the effect of the twins on $\rho_{xx}$ is reversed, similar to measurements reported for example in ref. [21]. Our voltage data collected using the various contacts on the crystal surface, show that for $\alpha=0°$ the vortices are progressively canalized by the dominant twin family for $B<B_{TB}$. The strong effect of twin-boundaries at $\alpha=0°$ is even more dramatic in Fig. 2 on the Hall conductivity, which is strongly reduced with respect to the Hall conductivity obtained at $\alpha=4°$. In fact, for $\alpha<1°$ twin-boundaries progressively "kill" the Hall conductivity below roughly 2.8T. Since $B_m \approx 2.1$ T, *this is an example where extended, strong pinning defects influence the Hall conductivity in the vortex-liquid phase*. For angles $\alpha>1°$, the twin-boundaries are less effective and the Hall conductivity can occur. For $\alpha=4°$, the Hall conductivity is not affected by twin-boundaries in the vortex-liquid phase, and deviates from the liquid behavior only at $B_m$.

The sharp change of the Hall conductivity observed at the vortex-lattice melting transition in Fig. 1 and Fig. 2 is not explained by the present microscopic theories, and represents the principal result of this paper. Most of the existing theoretical work about the Hall anomaly considers a single vortex moving in the presence of an applied transport current. The hydrodynamic and core Hall forces are obtained considering the momentum transfer from the moving vortex to the superfluid pairing state and the delocalized quasiparticles, and to the quasiparticles localized within the vortex core[3,7]. In particular, the sign of the Hall effect is ascribed to the difference of particle density at the vortex core and far outside the vortex core, i.e. details of the electronic band strucutre are indeed important. However, the possibility that the vortex-lattice melting transition may be related to changes of electronic degrees of freedom on the vortex background[13], or that microscopic fluctuations begin at the transition[14,15], is not considered in the Hall problem, nor are possible collective effects in the vortex-solid[26]. The data reported here show that the Hall conductivity deviates from the vortex-liquid behavior when the vortices "crystallize". By further decreasing the temperature or the field, the Hall conductivity increases in absolute value much faster than in the vortex-liquid phase, suggesting a change in the microscopic processes. Beyond the importance to the Hall problem, this finding is relevant to the issue of the nature of the paring process in high temperature superconductors[27] and to the question of the extent of the critical fluctuation region[28].

We thank G. Blatter and V. Geshkenbein for useful discussions. This work is supported by the Swiss National Science Foundation.

Figure captions:

FIG. 1. (a) The temperature dependence of the longitudinal resistivity $\rho_{xx}$ and the Hall resistivity $\rho_{xy}$ in a $YBa_2Cu_3O_{7-\delta}$ crystal at 2T, for various AC and DC current densities, as indicated for each curve as follow: ($j_{DC}$,$j_{AC}$) with both current densities in $A/cm^2$. The magnetic field is set at 4° from the c-axis in order to inhibit the effect of twin-boundaries. Inset: $\rho_{xx}$ versus T for various fields at 0 DC current and low AC current density. (b) The tangent of the Hall angle, $\tan\theta_H = \rho_{xy}/\rho_{xx}$, and the Hall conductivity, $\sigma_{xy} \approx \rho_{xy}/\rho_{xx}^2$, for low and high current densities as indicated. The dotted vertical line denotes the vortex-lattice melting transition at $T_m$. Inset: The Hall conductivity for various fields measured with a large DC current.

FIG. 2. The field dependence of the longitudinal resistivity $\rho_{xx}$ and the Hall conductivity, $\sigma_{xy} \approx \rho_{xy}/\rho_{xx}^2$, in the same $YBa_2Cu_3O_{7-\delta}$ crystal at 88 K, for various AC and DC current densities, as indicated for each curve as in Fig. 1. The magnetic field is at 4° and at 0° from the c-axis, showing the effect of twin-boundaries below about 3.4T. The dotted vertical line denotes the vortex-lattice melting transition at $B_m$. Inset: The Hall conductivity times the induction, $\sigma_{xy}B$, for various temperatures measured with a large DC current.

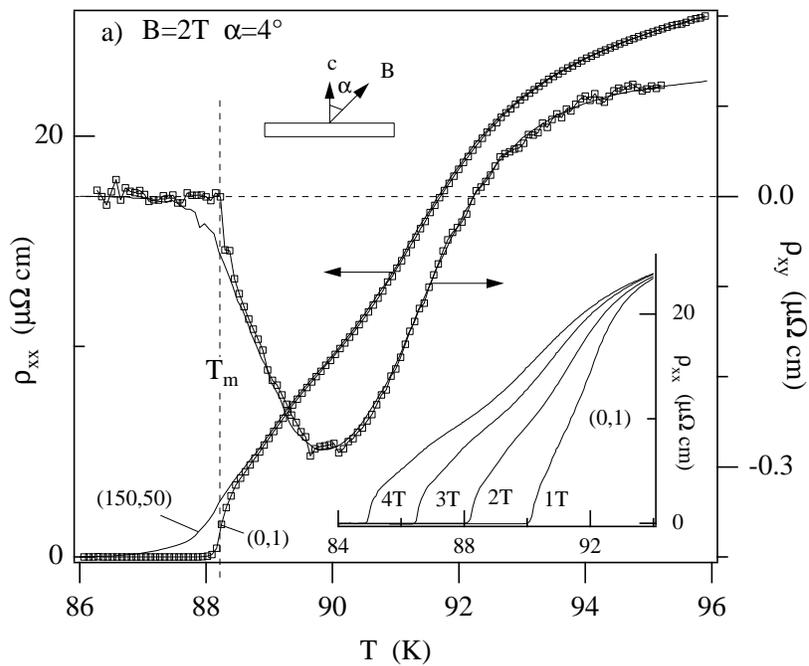

FIG. 1a)

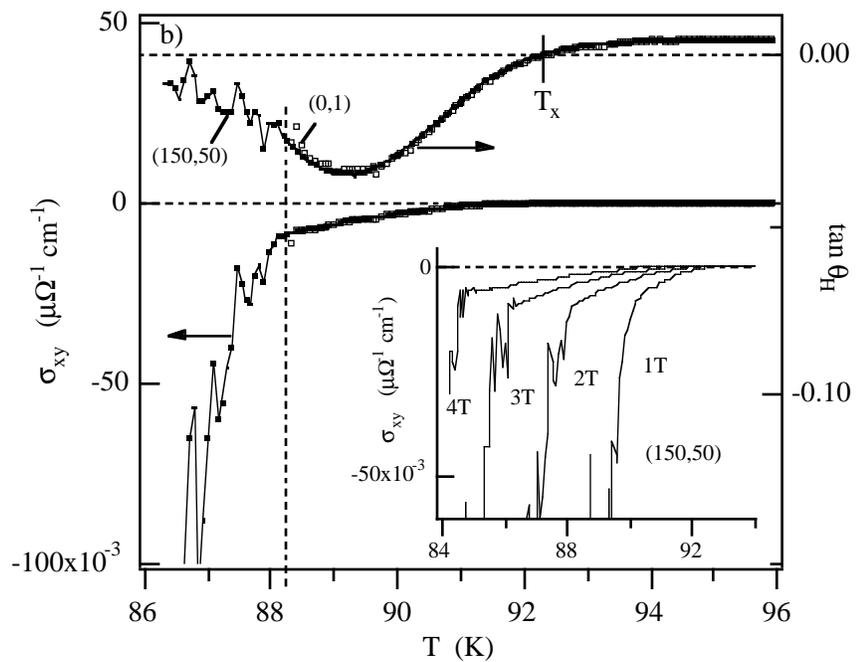

FIG. 1b)

FIG. 2)

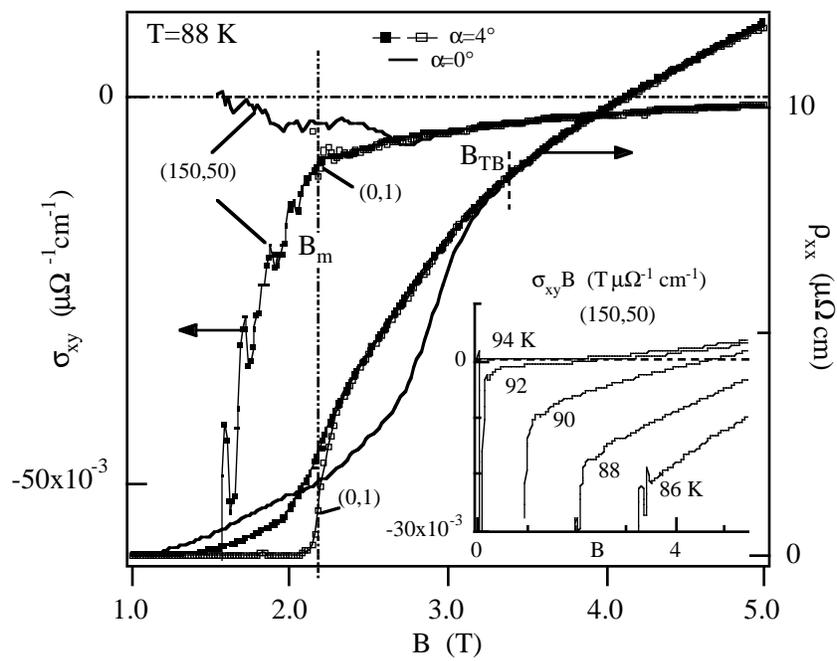